\documentclass[lettersize,journal]{IEEEtran}
\usepackage{amsmath,amsfonts}
\usepackage{algorithm}
\usepackage{array}
\usepackage{textcomp}
\usepackage{stfloats}
\usepackage{url}
\usepackage{verbatim}
\usepackage{graphicx}
\usepackage{cite}

\usepackage{subfigure}
\usepackage{algorithm,algpseudocode,multicol}
\usepackage{float}
\usepackage{soul} 
\usepackage{xcolor}
\usepackage{textcomp}
\usepackage{balance}
\usepackage{stfloats}
\usepackage{url}
\usepackage{verbatim}
\usepackage{threeparttable}
\usepackage{multirow}

\usepackage{amsmath,amssymb,amsfonts}
\usepackage{textcomp}
\usepackage{mathrsfs}
\usepackage{soul, color, xcolor} 

\newtheorem{thm}{Theorem}[section]
\newtheorem{lem}{Lemma}[section]

\newtheorem{defn}{Definition}[section]
\newtheorem{assm}{Assumption}[section]
\newtheorem{corol}{Corollary}[section]
\newtheorem{exam}{Example}[section]
\newtheorem{rem}{Remark}[section]

\usepackage{colortbl}

\usepackage{diagbox}

\begin{document}

\title{Cost-Aware Distributed Online Learning with Strict Rejection Behavior against Adversarial Agents}

\author{
Yuhan Suo, Runqi Chai, \emph{Senior Member, IEEE},
Senchun Chai, \emph{Senior Member, IEEE},\\
Xudong Zhao, \emph{Senior Member, IEEE},
Yuanqing Xia, \emph{Fellow, IEEE}
\thanks{
 Yuhan Suo, Runqi Chai, Senchun Chai,  and Yuanqing Xia  are with the School of Automation, 
Beijing Institute of Technology, Beijing 100081, China 
(e-mail: yuhan.suo@bit.edu.cn; chaisc97@bit.edu.cn; xia\_yuanqing@bit.edu.cn; r.chai@bit.edu.cn).
}
\thanks{
Yuhan Suo is also with Zhongguancun Academy, Beijing 100192, China.
}
\thanks{
Xudong Zhao is with the Key Laboratory of Intelligent
Control and Optimization for Industrial Equipment of Ministry of Education of Dalian University of Technology, 
Dalian 116024, China (e-mail: xdzhao-hit@gmail.com).
}
\thanks{
\emph{Corresponding author: Runqi Chai}.
}
}



\maketitle

\begin{abstract}
Distributed online learning in Internet of Things(IoT)-enabled multi-agent systems(MASs) is highly vulnerable to persistent adversarial interactions, particularly when malicious agents cannot be fully isolated during the transient learning stage. Existing resilient learning methods mainly focus on convergence preservation or malicious suppression, while the resulting evolution inefficiency caused by repeated corrective adaptation remains largely unexplored. To address this issue, this paper develops a cost-aware distributed online learning framework with a strict rejection behavior against adversarial agents. The proposed mechanism suppresses harmful assimilation of suspicious neighboring information and reveals a previously overlooked side effect, that is, the strict rejection may induce heterogeneous transient evolution among neighboring normal agents, leading to evolution desynchronization across the network. To mitigate this effect, a two-time-scale adaptive evolution regulation architecture is further developed, in which the outer layer dynamically adjusts the long-term evolution-rate schedule while the inner layer preserves robust online learning. Theoretical analysis establishes the dynamic tracking property of the outer-layer update and proves that the proposed regulation mechanism attenuates the propagation of strict-rejection-induced evolution desynchronization. Numerical simulations and a satellite-assisted IoT monitoring scenario demonstrate that the proposed method achieves robust and low-cost distributed online learning under persistent malicious interference.

\end{abstract}


\section{Introduction}
With the rapid advancement of the information era and the IoT, multi-agent systems have been increasingly deployed in various networked and security-sensitive applications, such as intelligent transportation systems \cite{dai2024marp}, sensor networks \cite{suo2025active}, and IoT-enabled distributed decision-making scenarios \cite{11153392}. In these systems, a group of autonomous agents interact and exchange information through local communication networks, enabling them to collaboratively accomplish global objectives that are difficult or inefficient for isolated agents to achieve.

Traditional interactive algorithms usually assume that all participating agents are benign and cooperative, and under this assumption they can guarantee convergence to a unified state with favorable convergence speed and robustness properties in benign settings \cite{amirkhani2022consensus}. Representative examples include average consensus algorithms \cite{de2022fundamental,10623393}, Laplacian-based protocols \cite{griparic2022consensus}, and robust fusion schemes designed to handle uncertainties such as noise and model perturbations \cite{liu2016robust}. These methods have laid an important foundation for distributed coordination and online adaptation. Nevertheless, their theoretical guarantees are typically established under idealized conditions in which the exchanged information is trustworthy or only affected by non-adversarial disturbances. In practical networked environments, especially those involving cyber-physical infrastructures or open communication platforms, such assumptions can be overly restrictive. Malicious agents may intentionally spread false information, manipulate local updates, or strategically reject the decisions of neighboring agents, thereby undermining both the correctness and the stability of the learning process \cite{shi2023secure}. From the perspective of information security, this means that the challenge is no longer limited to achieving consensus, but extends to maintaining reliable fusion and safe adaptation in the presence of adversarial behaviors.

To address these threats, recent studies have investigated a variety of resilient or secure distributed learning mechanisms. Existing approaches include detecting and isolating malicious agents according to abnormal signals or suspicious state patterns \cite{yue2023resilient}, assigning trust values or beliefs to modulate the influence of neighbors \cite{kayaalp2024social,cavorsi2023exploiting,b06a1e49b8a44ea69478c5e207c0c9ed}, exploiting redundancy in network information \cite{10750303}, using residual-based anomaly detection schemes \cite{zhou2022security,9889233}, and adopting moving target defense strategies to reduce attackers’ effectiveness \cite{9621221}. These studies have significantly improved the resilience of multi-agent coordination against malicious interference. 

However, as suggested in \cite{li2022robust,wu2023sequential}, the mitigation of adversarial influence in practical applications is rarely instantaneous; rather, it is typically a gradual process of attenuation and progressive isolation. This phenomenon generally arises for two reasons. First, to maximize the accuracy of identifying and eliminating malicious influence, the system often requires a certain period of information accumulation, behavioral observation, and cross-validation, so as to avoid misjudgment caused by insufficient evidence. For example, in distributed learning, multi-robot coordination, and sensor networks, a single abnormal state or a locally unusual residual is usually insufficient to justify the immediate isolation of an agent; instead, a reliable decision often requires integrating temporal data, neighborhood feedback, or evidence collected over multiple rounds of interaction. Second, due to the underlying task mechanism and operational constraints of the system itself, even when an agent is suspected of malicious behavior, its influence often cannot be completely removed at once. For example, in multi-robot formations, connected vehicle systems, and distributed satellite or UAV networks, a suspicious node may still be responsible for local sensing, communication relaying, or maintaining network topology. Abruptly disconnecting such a node may therefore disrupt task continuity, system observability, or network connectivity. As a result, in many real-world systems, the more practical strategy is not instantaneous isolation, but rather to progressively weaken the agent’s influence through trust degradation, interaction restriction, or conservative fusion until full isolation becomes feasible.

This transitional phase is particularly important from a security perspective, because normal agents may continue reacting to manipulated information while the attack is being mitigated. As a result, the system may suffer not only from degraded fusion accuracy, but also from repeated state corrections, amplified adjustment effort, and additional evolution cost. Regrettably, the vast majority of existing research tends to implicitly overlook this phenomenon, and therefore pay limited attention to how a distributed system should regulate its learning behavior online, using only locally available information and without relying on complete isolation, so as to preserve robustness while reducing unnecessary security-induced adaptation costs.

Compared with existing resilient distributed online learning methods
that primarily focus on convergence or malicious suppression
(e.g., \cite{cao2024discrete,10623393,10634212}),
this paper investigates the transient evolution dynamics induced by
adversarial interactions, particularly in scenarios where malicious
agents cannot be fully isolated during the online adaptation process.
Under persistent adversarial influence, normal agents may repeatedly
assimilate suspicious neighboring information and subsequently perform
corrective adaptation, thereby generating unnecessary transient
evolution burden and long-term learning inefficiency.

Existing trust-based interaction mechanisms
(e.g., \cite{b06a1e49b8a44ea69478c5e207c0c9ed})
mainly aim to suppress suspicious information at the fusion level.
However, the resulting transient evolution mismatch caused by repeated
interaction correction remains largely unexplored. In particular,
strict rejection of suspicious information may locally accelerate the
apparent evolution progress of directly affected agents, while
simultaneously inducing heterogeneous transient evolution among
neighboring normal agents. Such interaction-induced evolution mismatch
may further propagate through neighboring adaptation and amplify the
overall evolution burden of the distributed learning process.

Although several studies investigate minimum-cost formulations in
distributed optimization and learning
(e.g., \cite{guo2024multi,10804639,sun2025social}),
their analysis is mainly restricted to benign interaction settings and
does not characterize the evolution desynchronization induced by
adversarial interaction responses.
To address these issues, this paper develops a cost-aware distributed
online learning framework with strict rejection behavior and studies the
corresponding evolution desynchronization mechanism together with its
adaptive regulation architecture. Accordingly, the main contributions
are summarized as follows:
\begin{enumerate}

\item
A strict-rejection-based distributed online learning framework is developed for adversarial MASs. Different from existing resilient consensus methods that mainly focus on convergence or malicious isolation(e.g., \cite{cao2024discrete,10623393,10634212}), the proposed framework explicitly characterizes the evolution burden induced by adversarial interactions and reveals how strict rejection behavior may generate interaction-induced transient evolution mismatch among neighboring normal agents. To quantify this phenomenon, the notions of credible transient evolution gain and evolution desynchronization energy are introduced.

\item
A two-time-scale adaptive evolution regulation framework is proposed to suppress the propagation of strict-rejection-induced evolution desynchronization. In the proposed architecture, the fast inner-layer online adaptation and the slow outer-layer evolution-rate regulation are jointly coordinated. The resulting outer-layer update admits an equivalent formulation as a constrained online optimization problem with admissible-set projection and trust-region regulation, thereby providing a principled mechanism for regulating long-term evolution consistency under adversarial interactions.

\item
Theoretical guarantees are established for the proposed regulation framework. It is proved that the outer-layer update is well posed and achieves dynamic tracking of time-varying admissible evolution targets. Furthermore, a propagation attenuation theorem is established to show that the proposed regulation mechanism suppresses the network propagation of strict-rejection-induced evolution desynchronization. Finally, extensive simulations and a satellite-assisted IoT monitoring case study demonstrate that the proposed method achieves robust and low-cost distributed online adaptation under malicious interference.

\end{enumerate}

\section{Problem Formulation}

\subsection{Distributed Online Learning}
Consider an MASs network represented by an undirected graph $\mathcal{G}=(\mathcal{N},\mathcal{E})$, where $\mathcal{N}=\{1,2,\ldots,N\}$ is the node (agent) set and $\mathcal{E}\subseteq\mathcal{N}\times\mathcal{N}$ is the edge set. The neighbor set of agent $i$ is defined as $\mathcal{N}_i = \{\, j\in\mathcal{N} : (i,j)\in\mathcal{E}\,\}$.

At each time $k$, the goal of each agent $i$ is to minimize the local objective function $f_i(\cdot)$. It also needs to perform an information fusion operation based on the streaming information set $\xi_i(k) = \{x_j(k)\mid j\in\mathcal{N}_i\}$. 
Under normal interaction conditions, the fusion state is generated as
\begin{equation}\label{nominal_fusion}
x_i^f(k)
=
x_i(k)
+
\sum_{j\in\mathcal N_i}
\omega_{ij}(k)\big(x_j(k)-x_i(k)\big),
\end{equation}
where $\omega_{ij}(k)\ge0$ denotes the fusion weight associated with neighbor $j$.

The aforementioned process can be formulated as the following optimization problem:
\begin{equation}\label{local_objective}
    \min_{g_i(k)}\; f_i\big(x_i(k),g_i(k);\xi_i(k)\big)
\end{equation}
\vspace{-2mm}
\begin{equation}\label{follower_system}
		x_i(k+1)=A_i x_i^f(k)-M_i g_i(k),
	\end{equation}
where 
$x_i\left( k \right)\in \mathbb{R} ^{n}$, $x_i^f\left( k \right)\in \mathbb{R} ^{n}$,  $g_i(k)\in \mathbb{R} ^{m}$  represent the individual state, the fusion state of agent $i$, and the state evolution input at time step $k$, respectively. In addition, the system matrices $A_i$ and $M_i$ are matrices with appropriate dimensions.

In addition to normal agents, there are also some persistent malicious agents. They disregard the status of others and insist on spreading their own status \cite{cao2024discrete}.
	For such a malicious agent $i$, the fusion state coincides with its individual state, i.e., $x^f_i(k) = x_i(k)$. The state dynamics of a malicious agent still follow equation~(\ref{follower_system}), but its expected state $x^t_i$ is set outside the social norm region.


For a normal agent $i$, denote the fusion results in the absence and presence of malicious influence as $x_i^{f,u}$ and $x_i^{f,a}$, respectively. 
And $x_i^{f,a}$ is derived by taking into account the information $x_j^a(k)$ provided by a malicious neighbor $j$. Specifically, $x_j^a(k)$ can be written as 
\begin{equation}\label{malicious_inject}
    x_j^a(k) = x_j^{{ref}}(k) + z_j(k),
\end{equation}
where $x_j^{{ref}}(k)$ denotes the state of a non-malicious baseline system that shares the same weights and neighbor set as in the malicious scenario, and can thus be regarded as a \emph{virtual reference state} and $z_j(k)$ can be regarded as something similar to the malicious information mentioned in previous studies on malicious attacks.

\subsection{State Evolution Cost}
\begin{defn}\cite{zhou2014truncated}
Consider the state of agent $i$, which is updated online according to  (\ref{follower_system}). The optimization problem of any form of objective function $f_i(x)$ can be equivalently transformed into the problem of "the system state approaching the expected optimal state", i.e., $\min_x d(x_i,x_i^*)$. This paper aims to minimize the cost of this optimization process.
    Assume that $\{A_i, M_i\}$ is stabilizable. Define the state deviation as $\Tilde{x}_i(k)=x_i(k)-x_i^*$
        then, the online update sequence is determined by solving the following constrained optimization problem
\begin{equation}\label{optimal_control_problem}
\begin{aligned}
\min_{\{g_i(k)\}_{k=0}^{\infty}} \quad 
& J_i(g)
= \sum_{k=0}^{\infty}(1-\gamma_i)^{-k}
\left(
g_i^\top(k) R_i g_i(k)
\right) \\
\text{s.t.} \quad 
& \lim_{k\to\infty}\Tilde{x}_i(k)=0 \quad\text{and equation}  \quad (\ref{follower_system}).
\end{aligned}
\end{equation}
where $g_i$ is defined as $g_i(k)=K_i(k) \Tilde{x}_i(k) =(R_i+M_i^TP_i(k)M_i)^{-1}M_i^TP_i(k)A_i\Tilde{x}_i(k)$.
	and $P_i(k)$ is the iterative positive definite solution of the parametric discrete-time algebraic Riccati equation (PDARE) $\left(1-\gamma_i \right) P_i(k)=A_i^{\mathrm{T}} P_i(k+1) A_i
			-A_i^{\mathrm{T}} P_i(k+1) M_i
			\times\left(R_i+M_i^{\mathrm{T}} P_{i}(k+1) M_i\right)^{-1} M_i^{\mathrm{T}} P_i(k+1) A_i$.
	And for the case of infinite time period, forward iteration will eventually converge to the unique positive definite solution.
	Given the matrices $\{A_i, M_i,R_i\}$, the convergence rate of $K_i$ is closely related to the convergence rate of $P_i$. 
    \end{defn}

\subsection{Problem of Interest}

In distributed online learning over multi-agent systems, malicious
agents may remain temporarily unisolated during the transient
adaptation stage. Under persistent adversarial interactions, normal
agents may repeatedly assimilate suspicious neighboring information
through nominal fusion, thereby inducing cumulative transient state
deviation and unnecessary corrective evolution effort during the
subsequent online adaptation process.

The strict rejection behavior aims to suppress such harmful discrepancy
assimilation by reversing the interaction response to suspicious
neighboring discrepancies. However, strict rejection not only modifies
the information-fusion process, but also alters the transient evolution
trajectory of neighboring normal agents. In particular, directly
affected agents may exhibit externally accelerated transient evolution,
while neighboring normal agents continue evolving under heterogeneous
credible transient gains. Consequently, the resulting evolution
desynchronization may further propagate through neighboring adaptation
and amplify the long-term evolution burden of the distributed online
learning process.

This observation reveals a fundamental tension in resilient distributed
online learning: suppressing suspicious neighboring assimilation may
simultaneously induce heterogeneous transient evolution among normal
agents, thereby generating large-scale evolution desynchronization
across the network. Consequently, resilient online adaptation requires
not only adversarial interaction suppression, but also coordinated
regulation of the resulting evolution dynamics.

\section{Main Results}
This section develops the main results of the proposed framework. We first introduce the strict rejection behavior for suppressing suspicious neighboring interactions, and then characterize the evolution desynchronization it may induce among normal agents. Based on this observation, a two-time-scale adaptive evolution regulation mechanism is established, together with its dynamic tracking and propagation attenuation properties.

\subsection{Strict Rejection Behavior}

Consider the neighboring discrepancy induced by the nominal fusion
mechanism
\begin{equation}\label{neighbor_discrepancy}
\delta_i(k)
:=
\sum_{j\in\mathcal N_i}
\omega_{ij}(k)
\big(x_j(k)-x_i(k)\big).
\end{equation}

Under benign interactions, the neighboring discrepancy $\delta_i(k)$ usually assists the subsequent local online adaptation by reducing the disagreement among neighboring agents. However, under persistent adversarial interactions, suspicious neighboring discrepancies may instead increase the evolution burden associated with the subsequent local update process.

To characterize this effect, define the induced evolution burden generated by the neighboring discrepancy as
\begin{equation}\label{evolution_burden}
\mathcal E_i\big(\delta_i(k)\big)
:=
\left\|
M_i^\dagger
\big(
A_i(x_i(k)+\delta_i(k))
-
x_i^*
\big)
\right\|^2,
\end{equation}
where $x_i^*$ denotes the desirable local evolution target and $M_i^\dagger$ is the Moore--Penrose inverse of $M_i$.

For nominal cooperative interaction, attractive fusion responses typically decrease the induced burden $\mathcal E_i\big(\delta_i(k)\big)$ by moving neighboring states toward a common evolution direction. Nevertheless, under adversarial neighboring discrepancy, this monotonic relationship may no longer hold. In particular, repeatedly assimilating suspicious neighboring discrepancies may enlarge the subsequent corrective evolution effort and thereby amplify the transient evolution burden.

A straightforward remedy is to directly reverse the neighboring interaction once the discrepancy becomes suspicious. However,
excessively strong repulsive interaction may also become undesirable. Indeed, strong rejection responses may create interaction-induced
shortcut effects, where neighboring interaction itself artificially moves the state toward the desirable region without sufficient intrinsic
evolution effort. Although such displacement may reduce the apparent instantaneous evolution cost, the resulting improvement is induced by
external interaction rather than by the local online optimization process, thereby potentially leading to over-aggressive transient adaptation behavior.


Consequently, the interactive response should simultaneously suppress harmful differential assimilation and avoid excessive interaction-induced shift.

To achieve this balance, we introduce a generalized interaction-response operator
\begin{equation}\label{generalized_fusion}
x_i^f(k)
=
x_i(k)
+
\Phi_i\big(\delta_i(k)\big),
\end{equation}
where
$\Phi_i:\mathbb R^n\rightarrow\mathbb R^n$
modulates the neighboring interaction response according to the induced evolution burden.

The interaction-response operator is determined through the following regularized local selection problem:
\begin{equation}\label{Phi_optimization}
\Phi_i^\star
\in
\arg\min_{\Phi_i\in\mathcal A_i}
J_i^{\rm int}\big(\Phi_i\big),
\end{equation}
where the interaction-response objective is defined as
\begin{equation}\label{interaction_cost}
J_i^{\rm int}\big(\Phi_i\big)
:=
\mathcal E_i\big(\Phi_i(\delta_i(k))\big)
+
\lambda_i
\left\|
\Phi_i(\delta_i(k))
-
\delta_i(k)
\right\|^2,
\end{equation}
with $\lambda_i>0$ denoting the regularization coefficient and $\mathcal A_i$ representing the admissible interaction-response class.

The first term minimizes the predicted evolution burden after fusion, whereas the second term penalizes excessive deviation from nominal
cooperative interaction, thereby suppressing overly aggressive interaction-induced displacement.

Moreover, the admissible interaction-response class $\mathcal A_i$ is required to satisfy the discrepancy-response geometry
\begin{equation}\label{geometry_condition}
\left\langle
\Phi_i(\delta),\delta
\right\rangle
\begin{cases}
>0,
& \delta\in\mathcal D_i^{\rm a},
\\[4pt]
\le0,
& \delta\in\mathcal D_i^{\rm s},
\end{cases}
\end{equation}
where $\mathcal D_i^{\rm a}$ and $\mathcal D_i^{\rm s}$ denote the admissible and suspicious discrepancy regions, respectively.

Condition \eqref{geometry_condition} implies that admissible neighboring discrepancies preserve cooperative attractive interaction, whereas
suspicious discrepancies no longer induce positive assimilation energy. Consequently, the strict rejection behavior naturally emerges as the
interaction-response regime satisfying 
\begin{equation}\label{strict_rejection_condition}
\left\langle
\Phi_i(\delta),\delta
\right\rangle
<0,
\qquad
\delta\in\mathcal D_i^{\rm s}.
\end{equation}

An admissible realization is given below.

\begin{exam}
Consider the threshold-responsive interaction operator
\begin{equation}
\Phi_i(\delta)
=
\begin{cases}
\;\;\,\delta,
& \|\delta\|\le\theta_i,
\\[4pt]
-\alpha_i\delta,
& \|\delta\|>\theta_i,
\end{cases}
\end{equation}
where $\theta_i>0$ is the admissible discrepancy threshold and $\alpha_i\in(0,1]$ determines the rejection strength. The special case $\alpha_i=1$ recovers the hard strict rejection rule.
\end{exam}

The above analysis shows that strict rejection behavior is not introduced as a heuristic sign-reversal mechanism. Instead, it emerges as the interaction-response regime minimizing the compromise between harmful discrepancy assimilation and excessive interaction-induced displacement under adversarial neighboring interactions.

\begin{rem}
This strict rejection behavior has significant practical implications. For example, in social networks, individuals must strictly reject extremist or hate speech \cite{castano2021internet}; in public safety and finance, fraudulent rumors cannot be assimilated \cite{hernandez2024financial}. In these cases, the adversarial fusion rule provides a mechanism to encode this strict rejection behavior.
\end{rem}

\subsection{Evolution Desynchronization Induced by Strict Rejection}

The strict rejection mechanism developed in the previous subsection suppresses the assimilation of suspicious neighboring discrepancies by modifying the interaction response. However, such locally repulsive interaction may also introduce heterogeneous transient evolution among neighboring normal agents. This phenomenon becomes particularly relevant when some normal agents are directly exposed to malicious discrepancies, whereas their neighboring normal agents only receive attenuated indirect influence through the network fusion process.

For each normal agent $i$, define the interaction-induced displacement
as $\zeta_i(k)
:=
\Phi_i\big(\delta_i(k)\big)$.
Then the fusion state satisfies
$x_i^f(k)
=
x_i(k)
+
\zeta_i(k)$.
Substituting it into the state dynamics gives
\begin{equation}\label{state_with_zeta}
x_i(k+1)
=
A_i x_i(k)
+
A_i\zeta_i(k)
-
M_i g_i(k),
\end{equation}
which shows that the evolution process contains two fundamentally different components. The term $M_i g_i(k)$ represents the intrinsic local evolution effort generated by online adaptation, whereas $A_i\zeta_i(k)$ represents the external state displacement induced by neighboring interaction.

To quantify the influence of interaction-induced displacement, define the external evolution ratio
\begin{equation}\label{external_ratio}
\chi_i(k)
:=
\frac{\|A_i\zeta_i(k)\|^2}
{\|A_i\zeta_i(k)\|^2+\|M_i g_i(k)\|^2+\varepsilon_i},
\end{equation}
where $\varepsilon_i>0$ is a small constant introduced to avoid degeneracy. A larger value of $\chi_i(k)$ indicates that the apparent state evolution of agent $i$ is more strongly dominated by external interaction displacement rather than intrinsic local adaptation.

Next, define the transient evolution gain generated by the neighboring interaction response as
\begin{equation}\label{local_gain}
\Delta_i(k)
:=
\mathcal E_i(0)
-
\mathcal E_i\big(\Phi_i(\delta_i(k))\big),
\end{equation}
where $\mathcal E_i(\cdot)$ is the induced evolution burden defined in \eqref{evolution_burden}. The quantity $\Delta_i(k)$ measures the instantaneous reduction of the predicted evolution burden generated by the interaction-response operator.

Since part of the apparent evolution gain may originate from the external interaction displacement $A_i\zeta_i(k)$ rather than from the intrinsic local adaptation process, we define the credible transient evolution gain as
\begin{equation}\label{credible_gain}
\widehat{\Delta}_i(k)
:=
\big(1-\chi_i(k)\big)\Delta_i(k).
\end{equation}
The factor $1-\chi_i(k)$ discounts the portion of transient evolution gain dominated by neighboring interaction displacement.

For neighboring normal agents, define the local evolution desynchronization energy
\begin{equation}\label{local_desync_energy}
\widehat{\mathcal S}_i(k)
:=
\sum_{l\in\mathcal N_i\cap\mathcal N_u}
a_{il}
\big(
\widehat{\Delta}_i(k)
-
\widehat{\Delta}_l(k)
\big)^2,
\end{equation}
where $a_{il}\ge0$ denotes the neighboring coupling weight between normal agents. Correspondingly, define the global evolution desynchronization energy as
\begin{equation}\label{global_desync_energy}
\widehat{\mathcal S}(k)
:=
\frac{1}{2}
\sum_{i\in\mathcal N_u}
\sum_{l\in\mathcal N_i\cap\mathcal N_u}
a_{il}
\big(
\widehat{\Delta}_i(k)
-
\widehat{\Delta}_l(k)
\big)^2.
\end{equation}

The quantity $\widehat{\mathcal S}(k)$ measures the mismatch of credible transient evolution gains among neighboring normal agents. Large values of $\widehat{\mathcal S}(k)$ indicate that neighboring agents are evolving under significantly different transient adaptation responses.

The following result shows that strict rejection may induce nonzero evolution desynchronization among neighboring normal agents.

\begin{lem}\label{lem:strict_rejection_desync}
Consider two neighboring normal agents $A$ and $B$ with similar nominal evolution objectives. Suppose that agent $A$ is directly affected by a malicious neighboring discrepancy and undergoes strict rejection, whereas agent $B$ only receives attenuated indirect influence through neighboring propagation. Hence, the transient evolution of agent $A$ contains a non-negligible external-interaction component.
\end{lem}

\begin{IEEEproof}
Since agents $A$ and $B$ are neighboring normal agents, the global desynchronization energy $\widehat{\mathcal S}(k)$ contains the
quadratic term
\begin{equation}\label{AB_term}
a_{AB}
\big(
\widehat{\Delta}_A(k)
-
\widehat{\Delta}_B(k)
\big)^2.
\end{equation}
Using \eqref{credible_gain}, one has
$\widehat{\Delta}_A(k)
=
\big(1-\chi_A(k)\big)\Delta_A(k)$,
and
$\widehat{\Delta}_B(k)
=
\big(1-\chi_B(k)\big)\Delta_B(k)$.
Under conditions $\chi_A(k)
>
\chi_B(k)$ and $\Delta_A(k)
>
\Delta_B(k)$, the two credible transient evolution gains are generally mismatched except for the degenerate cancellation case. In particular,
$\widehat{\Delta}_A(k)
\neq
\widehat{\Delta}_B(k)$,
which implies that the quadratic term in \eqref{AB_term} is strictly positive. Therefore,
$
\widehat{\mathcal S}(k)>0.
$

Next, using \eqref{external_ratio} and
$\|A_A\zeta_A(k)\|
\ge
c_A\|M_Ag_A(k)\|$, $
c_A>0$, one obtains
$
\chi_A(k)
=
\frac{\|A_A\zeta_A(k)\|^2}
{\|A_A\zeta_A(k)\|^2+\|M_Ag_A(k)\|^2+\varepsilon_A}
\ge
\frac{
c_A^2\|M_Ag_A(k)\|^2
}{
(1+c_A^2)\|M_Ag_A(k)\|^2
+
\varepsilon_A
}$.
This completes the proof.
\end{IEEEproof}

Lemma~\ref{lem:strict_rejection_desync} indicates that strict rejection, while suppressing suspicious discrepancy assimilation, may induce heterogeneous transient evolution among neighboring normal agents. In particular, a directly affected agent can exhibit an excessively large credible transient gain relative to its neighbors. This observation suggests that the induced desynchronization penalty should favor reducing the evolution-rate schedule of such externally accelerated agents.

\begin{lem}[Local rate-reduction direction induced by evolution desynchronization]
\label{lem:penalty_gradient}
Suppose that the period-level credible transient gain
$\widehat{\Delta}_i^s$ is differentiable with respect to the
evolution-rate schedule $\gamma_i^s$. Consider the local
desynchronization penalty
\begin{equation}\label{desync_penalty}
\mathcal P_i^s(\gamma_i^s)
:=
\rho_s
\sum_{l\in\mathcal N_i\cap\mathcal N_u}
a_{il}
\big(
\widehat{\Delta}_i^s
-
\widehat{\Delta}_l^s
\big)^2,
\end{equation}
where $\rho_s>0$. In the local sensitivity analysis of agent $i$,
the neighboring credible gains $\widehat{\Delta}_l^s$ are treated as
fixed. Then
\begin{equation}\label{penalty_gradient}
\nabla_{\gamma_i^s}
\mathcal P_i^s(\gamma_i^s)
=
2\rho_s
\sum_{l\in\mathcal N_i\cap\mathcal N_u}
a_{il}
\big(
\widehat{\Delta}_i^s
-
\widehat{\Delta}_l^s
\big)
\nabla_{\gamma_i^s}
\widehat{\Delta}_i^s .
\end{equation}

Assume further that there exist constants
$\nu_i^s>0$ and $m_i^s>0$ such that
\begin{equation}\label{credible_gain_gap}
\widehat{\Delta}_i^s
-
\widehat{\Delta}_l^s
\ge
\nu_i^s,
\qquad
\forall l\in\mathcal N_i\cap\mathcal N_u
\textnormal{ with } a_{il}>0,
\end{equation}
and
\begin{equation}\label{positive_gain_sensitivity}
\nabla_{\gamma_i^s}
\widehat{\Delta}_i^s
\succeq
m_i^s\mathbf 1_T .
\end{equation}
If
$
\sum_{l\in\mathcal N_i\cap\mathcal N_u}a_{il}>0,
$
then
\begin{equation}\label{positive_penalty_gradient}
\nabla_{\gamma_i^s}
\mathcal P_i^s(\gamma_i^s)
\succeq
2\rho_s\nu_i^s m_i^s
\left(
\sum_{l\in\mathcal N_i\cap\mathcal N_u}
a_{il}
\right)
\mathbf 1_T
\succ0 .
\end{equation}
Consequently, any nonzero componentwise rate-reduction direction
$\vartheta_i^s\preceq0$ satisfies
\begin{equation}\label{directional_descent_penalty}
\left\langle
\nabla_{\gamma_i^s}
\mathcal P_i^s(\gamma_i^s),
\vartheta_i^s
\right\rangle
<0.
\end{equation}
Hence, when agent $i$ exhibits an excessive credible transient gain relative to its neighboring normal agents, reducing its evolution-rate schedule locally decreases the desynchronization penalty to first order.
\end{lem}

\begin{IEEEproof}
Differentiating \eqref{desync_penalty} with respect to $\gamma_i^s$, while treating the neighboring credible gains $\widehat{\Delta}_l^s$ as locally fixed, yields
\begin{equation}
\nabla_{\gamma_i^s}
\mathcal P_i^s(\gamma_i^s)
=
2\rho_s
\sum_{l\in\mathcal N_i\cap\mathcal N_u}
a_{il}
\big(
\widehat{\Delta}_i^s
-
\widehat{\Delta}_l^s
\big)
\nabla_{\gamma_i^s}
\widehat{\Delta}_i^s ,
\end{equation}
which proves \eqref{penalty_gradient}.

Under \eqref{credible_gain_gap} and
\eqref{positive_gain_sensitivity}, each weighted summand satisfies
\begin{equation}
a_{il}
\big(
\widehat{\Delta}_i^s
-
\widehat{\Delta}_l^s
\big)
\nabla_{\gamma_i^s}
\widehat{\Delta}_i^s
\succeq
a_{il}\nu_i^s m_i^s\mathbf 1_T .
\end{equation}
Summing over
$l\in\mathcal N_i\cap\mathcal N_u$
gives
$
\nabla_{\gamma_i^s}
\mathcal P_i^s(\gamma_i^s)
\succeq
2\rho_s\nu_i^s m_i^s
\left(
\sum_{l\in\mathcal N_i\cap\mathcal N_u}
a_{il}
\right)
\mathbf 1_T$.
Since $\rho_s>0$, $\nu_i^s>0$, $m_i^s>0$, and
$\sum_{l\in\mathcal N_i\cap\mathcal N_u}a_{il}>0$,
\eqref{positive_penalty_gradient} follows.

Finally, let $\vartheta_i^s\preceq0$ be any nonzero componentwise
rate-reduction direction. From
\eqref{positive_penalty_gradient},
\begin{equation}
\left\langle
\nabla_{\gamma_i^s}
\mathcal P_i^s(\gamma_i^s),
\vartheta_i^s
\right\rangle
\le
2\rho_s\nu_i^s m_i^s
\left(
\sum_{l\in\mathcal N_i\cap\mathcal N_u}
a_{il}
\right)
\mathbf 1_T^\top \vartheta_i^s
<0,
\end{equation}
because $\mathbf 1_T^\top \vartheta_i^s<0$. This proves
\eqref{directional_descent_penalty} and completes the proof.
\end{IEEEproof}

The above result provides a local variational justification for the subsequent outer-layer evolution-rate regulation. It shows that rejection-induced desynchronization naturally gives rise to a rate-reduction tendency for directly accelerated agents. Therefore, resilient distributed online learning requires a coordinated evolution-rate regulation mechanism operating beyond the instantaneous fusion layer.

\subsection{Two-Time-Scale Adaptive Evolution Regulation Framework}

The previous subsection shows that strict rejection may induce neighboring evolution desynchronization among normal agents. In particular, directly affected agents may exhibit externally accelerated transient evolution, whereas neighboring normal agents evolve under
significantly different credible transient gains. Consequently, additional regulation mechanisms are required to suppress excessive desynchronization propagation over long-term adaptation.

To mitigate such transient amplification effects, we introduce a two-time-scale adaptive evolution regulation framework. The key idea is to separate the fast inner-layer state evolution from the slow outer-layer evolution-rate adaptation. At the fast time scale, each
normal agent evolves under a fixed periodic evolution-rate schedule within one outer period. At the slow time scale, the schedule is updated only once per outer period according to the neighboring evolution desynchronization level.

For each normal agent $i$, define the periodic evolution-rate schedule
$\gamma_i^s
:=
\mathrm{col}
\big(
\gamma_i^{(1,s)},
\ldots,
\gamma_i^{(T,s)}
\big)
\in\mathbb R^T$,
where $s$ denotes the outer-period index, $T$ is the period length, and $\tau\in\{1,\ldots,T\}$ indexes the intra-period slots.

The role of the outer layer is not to modify the state directly, but to adaptively regulate the long-term evolution-rate schedule in order to suppress excessive desynchronization propagation while preserving periodic consistency of the inner-layer dynamics.

To preserve admissibility of the evolution process, define the time-varying admissible set
$\Gamma_i^s
:=
\left\{
u\in\mathbb R^T:
\underline f_i^s
\le
u
\le
\overline f_i^s
\right\}$,
where
$
\underline f_i^s
=
\mathrm{col}
\big(
f_{i,\min}^{(s)}(1),
\ldots,
f_{i,\min}^{(s)}(T)
\big)
$
and
$
\overline f_i^s
=
\mathrm{col}
\big(
f_{i,\max}^{(s)}(1),
\ldots,
f_{i,\max}^{(s)}(T)
\big)
$
denote the componentwise admissible lower and upper bounds, respectively.

At the beginning of outer period $s$, each agent has access to an outer-layer information set $\mathcal I_i^s$, which may contain warning signals, historical schedules, neighboring desynchronization levels, or other exogenous indicators. Based on this information, a reference evolution scheduling scheme is generated via a pre-defined reference selection operator, that is, 
\begin{equation}
    \gamma_{i,\mathrm{ref}}^s
=
\mathcal R_i(\mathcal I_i^s).
\end{equation}

The reference schedule is interpreted as a desired slow-time-scale evolution profile that suppresses excessive neighboring evolution mismatch. In particular, agents exhibiting large desynchronization energy are encouraged to reduce their subsequent evolution rate, whereas agents with small neighboring mismatch preserve their nominal evolution schedule.

To synthesize the updated schedule, consider the outer-layer objective
\begin{equation}
F_i^s(u)
=
\frac{\lambda}{2}
\|u-\gamma_{i,\mathrm{ref}}^s\|^2
+
I_{\mathbb B_\infty(v_i^s,\delta)}(u),
\label{outer_objective}
\end{equation}
where $\lambda\in(0,1]$ is the target-pull coefficient,
$\delta>0$ is the trust-region radius, and
$
\mathbb B_\infty(v_i^s,\delta)
:=
\left\{
u\in\mathbb R^T:
\|u-v_i^s\|_\infty\le\delta
\right\}$.
Moreover,
$I_{\mathbb B_\infty(v_i^s,\delta)}(\cdot)$ denotes the indicator
function of the trust region.

The first term in \eqref{outer_objective} drives the schedule toward the desired reference profile, whereas the trust-region constraint prevents excessively aggressive inter-period variation caused by transient neighboring desynchronization.

Using the Euclidean regularizer
$
D_i(u,v_i^s)=\|u-v_i^s\|^2
$
and $\rho:=1-\lambda$, the outer-layer schedule synthesis problem is formulated as
\begin{equation}\label{outer_problem}
\begin{aligned}
\gamma_i^s
=
\arg\min_{u\in\mathbb R^T}
\quad
&
\frac{1-\lambda}{2}\|u-v_i^s\|^2
+
\frac{\lambda}{2}
\|u-\gamma_{i,\mathrm{ref}}^s\|^2
\\
\textnormal{s.t.}\quad
&
\|u-v_i^s\|_\infty\le\delta,
\\
&
u\in\Gamma_i^s.
\end{aligned}
\end{equation}

Problem \eqref{outer_problem} admits a natural dynamic-tracking interpretation. If the reference schedule remains unchanged, then the outer layer reduces to a static admissible regulation problem. Otherwise, the outer layer tracks a time-varying admissible evolution
profile generated by neighboring desynchronization feedback.

Since the objective in \eqref{outer_problem} is strongly convex and the feasible set is closed and convex, the optimizer exists uniquely. The corresponding unconstrained minimizer is
\begin{equation}
z_i^s
=
(1-\lambda)v_i^s
+
\lambda\gamma_{i,\mathrm{ref}}^s,
\end{equation}
where $v_i^s
=\gamma_i^{s-1}$, with initialization $v_i^0=\gamma_i^0$.

To explicitly enforce the trust-region constraint, define the componentwise clipping operator
\begin{equation}\label{clip_op}
\mathcal T_\delta(z;v)
:=
v
+
\mathrm{clip}
\big(
z-v,
-\delta\mathbf 1,
\delta\mathbf 1
\big),
\end{equation}
where $\mathbf1$ denotes the all-ones vector and
$\mathrm{clip}(\cdot)$ acts componentwise.

Then the optimizer of \eqref{outer_problem} is explicitly given by
\begin{equation}\label{outer_update}
\gamma_i^s
=
\Pi_{\Gamma_i^s}
\Big(
\mathcal T_\delta(z_i^s;v_i^s)
\Big),
\end{equation}
where $\Pi_{\Gamma_i^s}$ denotes the Euclidean projection onto the
admissible set $\Gamma_i^s$.

The update law \eqref{outer_update} admits a clear geometric interpretation. The previous schedule $v_i^s=\gamma_i^{s-1}$ provides a temporal anchor, the reference schedule provides the desired evolution direction, the trust-region clipping suppresses abrupt schedule variation, and the projection operator guarantees admissibility.

The following lemma summarizes several basic properties of the proposed outer-layer update.

\begin{lem}\label{lem:outer_basic}
Suppose that, for each $i\in\mathcal N_u$ and each $s\ge1$, the set
$\Gamma_i^s$ is nonempty, closed, and convex. Let
$v_i^s:=\gamma_i^{s-1}$. Then the update \eqref{outer_update} is well
defined and satisfies
$\gamma_i^s\in\Gamma_i^s$, $
\forall s\ge1$.
Moreover,
\begin{equation}\label{outer_update_deviation_bound}
\|\gamma_i^s-v_i^s\|
\le
\sqrt T\,\delta
+
\mathrm{dist}(v_i^s,\Gamma_i^s).
\end{equation}
\end{lem}

\begin{IEEEproof}
Since $\Gamma_i^s$ is nonempty, closed, and convex, the Euclidean projection $\Pi_{\Gamma_i^s}$ is single valued. Hence, \eqref{outer_update} is well defined, and
$\gamma_i^s\in\Gamma_i^s$ follows directly from the definition of the projection.

By construction of the clipping operator, each component of
$\mathcal T_\delta(z_i^s;v_i^s)-v_i^s$ belongs to
$[-\delta,\delta]$. Therefore,
$\big\|
\mathcal T_\delta(z_i^s;v_i^s)-v_i^s
\big\|
\le
\sqrt T\,\delta$.
Using the nonexpansiveness of the Euclidean projection, one obtains
$\big\|
\gamma_i^s
-
\Pi_{\Gamma_i^s}(v_i^s)
\big\|
\le
\big\|
\mathcal T_\delta(z_i^s;v_i^s)-v_i^s
\big\|$.
Hence,
$\|\gamma_i^s-v_i^s\|
\le
\big\|
\gamma_i^s-\Pi_{\Gamma_i^s}(v_i^s)
\big\|
+
\mathrm{dist}(v_i^s,\Gamma_i^s)
\le
\sqrt T\,\delta
+
\mathrm{dist}(v_i^s,\Gamma_i^s)$,
which proves \eqref{outer_update_deviation_bound}. This completes the
proof.
\end{IEEEproof}

The above properties imply that the proposed outer-layer regulation is feasible by construction and suppresses excessively abrupt schedule variation caused by neighboring evolution desynchronization.

\subsection{Dynamic Tracking and Propagation Attenuation}

The previous subsection constructs the outer-layer schedule update on time-varying admissible sets. We now show that this update tracks the admissible reference schedule and suppresses the propagation of strict-rejection-induced evolution desynchronization.

For each outer period $s$, define
the projected admissible target
\begin{equation}\label{eq:projected_target}
\bar\gamma_i^s
:=
\Pi_{\Gamma_i^s}
\big(
\gamma_{i,\mathrm{ref}}^s
\big)
\end{equation}
and the target drift
\begin{equation}\label{eq:target_drift}
D_i^s
:=
\left\|
\bar\gamma_i^s-\bar\gamma_i^{s-1}
\right\|.
\end{equation}
The anchor tracking error and the actual schedule error are defined as
\begin{equation}\label{eq:outer_errors}
e_i^s:=v_i^s-\bar\gamma_i^s,
\qquad
\tilde e_i^s:=\gamma_i^s-\bar\gamma_i^s.
\end{equation}


\begin{thm}[Tracking of drifting admissible targets]
\label{thm:outer_tracking}
Suppose that the clipping operator in \eqref{...} is inactive along the generated trajectory, so that
$\mathcal T_i^s(u)
=
\Pi_{\Gamma_i^s}
\big(
(1-\lambda)u+\lambda\gamma_{i,\mathrm{ref}}^s
\big)$ at the generated anchor point. Then,  for every $i\in\mathcal N_u$ and $s\ge1$,
\begin{equation}\label{eq:anchor_tracking}
\|e_i^s\|
\le
q_i\|e_i^{s-1}\|+D_i^s,
\end{equation}
where
$q_i:=1-\lambda\in[0,1)$.
Moreover, the actual schedule error satisfies
$\|\tilde e_i^s\|
\le
(1-\lambda)\|e_i^s\|$.
Consequently, if $\sup_s D_i^s\le \bar D_i<\infty$, then
\begin{equation}\label{eq:outer_limsup_e}
\limsup_{s\to\infty}\|e_i^s\|
\le
\frac{\bar D_i}{\lambda},\quad
\limsup_{s\to\infty}\|\tilde e_i^s\|
\le
\frac{(1-\lambda)\bar D_i}{\lambda}.
\end{equation}
If $D_i^s\to0$, then $e_i^s\to0$ and $\tilde e_i^s\to0$.
\end{thm}

\begin{IEEEproof}
For each fixed $s$, the projected target satisfies
$\mathcal T_i^s(\bar\gamma_i^s)=\bar\gamma_i^s$. Since the Euclidean
projection is nonexpansive, $\mathcal T_i^s$ is $(1-\lambda)$-Lipschitz.
Thus,
\begin{equation}\label{eq:schedule_error_proof}
\|\tilde e_i^s\|
=
\|\mathcal T_i^s(v_i^s)-\mathcal T_i^s(\bar\gamma_i^s)\|
\le
(1-\lambda)\|e_i^s\|.
\end{equation}
Using $v_i^s=\gamma_i^{s-1}$, one has
\begin{equation}\label{eq:error_decomposition}
e_i^s
=
\tilde e_i^{s-1}
+
(\bar\gamma_i^{s-1}-\bar\gamma_i^s).
\end{equation}
Combining \eqref{eq:schedule_error_proof} and
\eqref{eq:error_decomposition} gives
$\|e_i^s\|
\le
\big[1-\lambda\big]\|e_i^{s-1}\|+D_i^s$,
which proves \eqref{eq:anchor_tracking}. The remaining statements follow from the scalar comparison recursion $a_s\le q_i a_{s-1}+D_i^s$ with $q_i\in[0,1)$. This completes the proof.
\end{IEEEproof}
\begin{rem}
Theorem~\ref{thm:outer_tracking} characterizes the dynamic tracking behavior in the contractive regime, where the trust-region clipping does not saturate along the generated trajectory. When the clipping is active, the trust-region still guarantees bounded inter-period variation, but the contraction estimate in Theorem~\ref{thm:outer_tracking} may no
longer hold with the same factor.
\end{rem}

The tracking result also yields a dissipative estimate for the outer
tracking error, which will be useful for the subsequent network-level
attenuation analysis.

\begin{corol}\label{cor:outer_dissipation}
Based on Theorem~\ref{thm:outer_tracking}, for any $\varepsilon>0$
satisfying $(1+\varepsilon)q_i^2<1$, the tracking error satisfies
\begin{equation}\label{eq:outer_dissipation}
\|e_i^s\|^2-\|e_i^{s-1}\|^2
\le
-\mu_i\|e_i^{s-1}\|^2
+
c_i(D_i^s)^2,
\end{equation}
where
$
\mu_i:=1-(1+\varepsilon)q_i^2>0$, $
c_i:=1+\frac{1}{\varepsilon}$.
\end{corol}

\begin{IEEEproof}
From \eqref{eq:anchor_tracking},
$\|e_i^s\|^2
\le
(q_i\|e_i^{s-1}\|+D_i^s)^2$.
Applying Young's inequality gives
$\|e_i^s\|^2
\le
(1+\varepsilon)q_i^2\|e_i^{s-1}\|^2
+
\left(1+\frac{1}{\varepsilon}\right)(D_i^s)^2$.
Then, subtract $\|e_i^{s-1}\|^2$ from both sides. This completes the proof.
\end{IEEEproof}

Next, we formalize how the above outer-layer regulation suppresses the network propagation of strict-rejection-induced desynchronization. Recall that $\widehat{\Delta}_i^s$ denotes the period-averaged credible transient gain and $\widehat{\mathcal S}_i^s$ denotes the local desynchronization energy. Define the directly affected set as
$\mathcal N_{\rm dir}^s
:=
\left\{
i\in\mathcal N_u:
\exists k\in\mathcal K_s
\ \textnormal{s.t.}\
\left\langle
\Phi_i(\delta_i(k)),\delta_i(k)
\right\rangle<0
\right\}$.
For $h\ge0$, define the shortest-distance layer
$\mathcal L_h^s
:=
\left\{
i\in\mathcal N_u:
\operatorname{dist}_{\mathcal G_u}
(i,\mathcal N_{\rm dir}^s)=h
\right\}$,
which means that if a normal node is one-hop from one directly affected region and two-hop from another, it belongs to the smaller-distance layer.

For convenience, let
$\widehat{\Delta}_i^{s+1}(\gamma)$ denote the period-level credible transient gain of agent $i$ in outer period $s+1$ when the evolution-rate schedule is fixed at $\gamma$. Hence, $\widehat{\Delta}_i^{s+1}
=
\widehat{\Delta}_i^{s+1}(\gamma_i^{s+1})$.
\begin{assm}[Inter-layer attenuation]
\label{ass:inter_layer_attenuation}
There exists $\beta\in(0,1)$ such that, for every $h\ge1$ and
$i\in\mathcal L_h^s$,
$|
\widehat{\Delta}_i^{s+1}
|
\le
\beta
\sum_{j\in\mathcal N_i\cap\mathcal L_{h-1}^s}
a_{ij}
|
\widehat{\Delta}_j^{s+1}
|$.
\end{assm}

\begin{lem}[Source-layer contraction under effective outer-layer regulation]
\label{lem:source_layer_contraction}
For every directly affected agent $i\in\mathcal L_0^s$, suppose that the outer-layer update induces a nontrivial componentwise reduction of the evolution-rate schedule, namely,
\begin{equation}\label{effective_schedule_reduction}
\gamma_i^{s+1}
\preceq
\gamma_i^s
-
\underline{\kappa}_i^s\mathbf 1_T,
\qquad
\underline{\kappa}_i^s>0.
\end{equation}
Moreover, suppose that there exist constants $m_i^s>0$ and
$\omega_i^s\ge0$ such that
\begin{equation}\label{credible_gain_rate_monotonicity}
\nabla_{\gamma}
\widehat{\Delta}_i^{s+1}(\gamma)
\succeq
m_i^s\mathbf 1_T
\end{equation}
along the line segment joining $\gamma_i^s$ and $\gamma_i^{s+1}$, and
\begin{equation}\label{credible_gain_frozen_drift}
\widehat{\Delta}_i^{s+1}(\gamma_i^s)
-
\widehat{\Delta}_i^s
\le
\omega_i^s.
\end{equation}
If
\begin{equation}\label{net_source_regulation_margin}
0
<
T m_i^s\underline{\kappa}_i^s-\omega_i^s
<
\widehat{\Delta}_i^s,
\qquad
\widehat{\Delta}_i^{s+1}\ge0,
\end{equation}
then there exists $\sigma_i^s\in(0,1)$ such that $\widehat{\Delta}_i^{s+1}
\le
(1-\sigma_i^s)
\widehat{\Delta}_i^s$
\end{lem}

\begin{IEEEproof}
For each $i\in\mathcal L_0^s$, define
$
\Delta\gamma_i^s
:=
\gamma_i^{s+1}-\gamma_i^s
$.
By \eqref{effective_schedule_reduction}, $\Delta\gamma_i^s
\preceq
-\underline{\kappa}_i^s\mathbf 1_T$ can be obtained.
Using the mean-value integral representation of
$\widehat{\Delta}_i^{s+1}(\gamma)$ along the segment joining
$\gamma_i^s$ and $\gamma_i^{s+1}$ gives
\begin{small}
   \begin{align}
\widehat{\Delta}_i^{s+1}(\gamma_i^{s+1})
-
\widehat{\Delta}_i^{s+1}(\gamma_i^s)=
\int_0^1
\left\langle
\nabla_{\gamma}
\widehat{\Delta}_i^{s+1}
\big(
\gamma_i^s+\tau\Delta\gamma_i^s
\big),
\Delta\gamma_i^s
\right\rangle
d\tau .
\label{mean_value_integral}
\end{align} 
\end{small}
By \eqref{credible_gain_rate_monotonicity}, the integrand satisfies
$\left\langle
\nabla_{\gamma}
\widehat{\Delta}_i^{s+1}
\big(
\gamma_i^s+\tau\Delta\gamma_i^s
\big),
\Delta\gamma_i^s
\right\rangle
\le
-
T m_i^s\underline{\kappa}_i^s$.
Substituting this estimate into \eqref{mean_value_integral} yields
\begin{equation}\label{credible_gain_schedule_decrease}
\widehat{\Delta}_i^{s+1}(\gamma_i^{s+1})
-
\widehat{\Delta}_i^{s+1}(\gamma_i^s)
\le
-
T m_i^s\underline{\kappa}_i^s .
\end{equation}
Since
$
\widehat{\Delta}_i^{s+1}
=
\widehat{\Delta}_i^{s+1}(\gamma_i^{s+1}),
$
combining \eqref{credible_gain_schedule_decrease} with
\eqref{credible_gain_frozen_drift} gives
\begin{align}
\widehat{\Delta}_i^{s+1}
\le
\widehat{\Delta}_i^{s+1}(\gamma_i^s)
-
T m_i^s\underline{\kappa}_i^s
\le
\widehat{\Delta}_i^s
+
\omega_i^s
-
T m_i^s\underline{\kappa}_i^s .
\label{credible_gain_combined_bound}
\end{align}
Define $\sigma_i^s
:=
\frac{
T m_i^s\underline{\kappa}_i^s-\omega_i^s
}{
\widehat{\Delta}_i^s
}$
By \eqref{net_source_regulation_margin}, $\sigma_i^s\in(0,1)$.
Therefore, we have
$\widehat{\Delta}_i^{s+1}
\le
(1-\sigma_i^s)
\widehat{\Delta}_i^s$. This completes the proof.
\end{IEEEproof}

\begin{rem}
Lemma~\ref{lem:source_layer_contraction} does not prescribe a specific reference-generation rule $\gamma_{i,\mathrm{ref}}^s=\mathcal R_i(\mathcal I_i^s)$.
It only characterizes the net rate-reduction effect produced by the outer-layer regulation. For instance, \eqref{effective_schedule_reduction} holds whenever the projected
reference target is sufficiently smaller than $\gamma_i^s$ and the outer-layer tracking error does not offset this reduction.
\end{rem}

\begin{thm}[Propagation attenuation of rejection-induced evolution desynchronization]
\label{thm:propagation_attenuation}
Suppose that the source-layer contraction in Lemma~\ref{lem:source_layer_contraction} and the inter-layer attenuation condition in Assumption~\ref{ass:inter_layer_attenuation} hold. Let
$
\sigma_{\min}^s
:=
\min_{i\in\mathcal L_0^s}\sigma_i^s.
$
Then, for every $h\ge1$, the layer-wise credible-gain energy satisfies
\begin{equation}\label{propagation_attenuation_bound}
\mathcal R_h^{s+1}
\le
C_h^s
\beta^{2h}
(1-\sigma_{\min}^s)^2
\mathcal R_0^s,
\end{equation}
where $C_h^s>0$ depends only on the inter-layer coupling weights.
\end{thm}

\begin{IEEEproof}
From Lemma~\ref{lem:source_layer_contraction}, for every
$i\in\mathcal L_0^s$,
$\widehat{\Delta}_i^{s+1}
\le
(1-\sigma_i^s)
\widehat{\Delta}_i^s$.
Since $\widehat{\Delta}_i^{s+1}\ge0$ and $\widehat{\Delta}_i^s>0$ under
\eqref{net_source_regulation_margin}, one also has
$\left|
\widehat{\Delta}_i^{s+1}
\right|
\le
(1-\sigma_i^s)
\left|
\widehat{\Delta}_i^s
\right|$.
Summing over all $i\in\mathcal L_0^s$ gives
\begin{equation}\label{source_layer_energy_contraction}
\mathcal R_0^{s+1}
\le
(1-\sigma_{\min}^s)^2
\mathcal R_0^s.
\end{equation}

We next propagate the source-layer suppression across graph layers. For every $h\ge1$ and every $i\in\mathcal L_h^s$, Assumption~\ref{ass:inter_layer_attenuation} gives
$\left|
\widehat{\Delta}_i^{s+1}
\right|
\le
\beta
\sum_{j\in\mathcal N_i\cap\mathcal L_{h-1}^s}
a_{ij}
\left|
\widehat{\Delta}_j^{s+1}
\right|$.
Applying Cauchy's inequality yields
\begin{align}
\left|
\widehat{\Delta}_i^{s+1}
\right|^2
\le
\beta^2
\left(
\sum_{j\in\mathcal N_i\cap\mathcal L_{h-1}^s}
a_{ij}
\right)
\sum_{j\in\mathcal N_i\cap\mathcal L_{h-1}^s}
a_{ij}
\left|
\widehat{\Delta}_j^{s+1}
\right|^2 .
\label{layer_cauchy_bound}
\end{align}
Define
$\bar a_h^s
:=
\max_{i\in\mathcal L_h^s}
\sum_{j\in\mathcal N_i\cap\mathcal L_{h-1}^s}
a_{ij}$, $
\bar b_h^s
:=
\max_{j\in\mathcal L_{h-1}^s}
\sum_{i\in\mathcal L_h^s\cap\mathcal N_j}
a_{ij}$.
Summing \eqref{layer_cauchy_bound} over
$i\in\mathcal L_h^s$ gives
$\mathcal R_h^{s+1}
\le
\beta^2
\bar a_h^s
\bar b_h^s
\mathcal R_{h-1}^{s+1}$.
Applying this inequality recursively from layer $1$ to layer $h$ yields
$\mathcal R_h^{s+1}
\le
\left(
\prod_{q=1}^{h}
\bar a_q^s\bar b_q^s
\right)
\beta^{2h}
\mathcal R_0^{s+1}$.
Let
$C_h^s
:=
\prod_{q=1}^{h}
\bar a_q^s\bar b_q^s $.
Combining the above estimate with
\eqref{source_layer_energy_contraction} proves
\eqref{propagation_attenuation_bound}. This completes the proof.
\end{IEEEproof}

\section{Numerical Simulation}
In this section,  $10$ agents are considered, which form an undirected graph. Among them, $\mathcal{N}^u=\{1,2,3,4\}$ are normal agents, while the rest are malicious. And the neighbor sets $\mathcal{N}_i$ of each normal agent $i=1:4$ are $\mathcal{N}_1=\{2,3,4,5,6,10\}$, $\mathcal{N}_2=\{1,3,4,7,9,10\}$, $\mathcal{N}_3=\{1,2,7,8,10\}$, $\mathcal{N}_4=\{1,2,5,7,8,10\}$.
To mitigate the influence of malicious agents, normal agents progressively reduce their interaction weights with them based on inter-agent distances. Once a weight falls below the threshold, the  link is cut off. 
	

The state evolution process of each agent $i$ satisfies the equation (\ref{follower_system}),
where the system matrix
$A_i=[0.99\quad -0.01\quad 0;-0.01\quad 0.99\quad 0;0\quad 0\quad 0.99]+rand()\cdot diag([1,1,1] / 100)$,
$M_i=[0 \quad0.5;0.5 \quad0;-0.5 \quad 0]$, the pair $\{A_i,M_i\}$ is stabilizable, and $\|A_j-A_k\|$ is bounded for each agent $j,k$.
The initial state of each agent $i$ is $x_i(0)=[0;0;0]$, and $R_i=I_{2\times 2}$. 
The expected states of normal agents are distributed within a small sphere centered at $[3,3,3]^T$, while those of malicious agents are symmetrically distributed around the opposite. 

The reference schedule generation method employed in this simulation is based on peak clipping (as shown in Appendix B), which was configured based on the specific requirements of this simulation.
According to Algorithm~\ref{alg:peak_clipping} in Appendix A, there are four cases of the reference sequence of the function $f_i(k)$
, with the initial and terminal function periods being $[850,1175,1413,1500,1413,1175,850]\times10^{-4}$ and $0.085\mathbf{1}_{1\times 6}$, respectively.
Given that the maximum variation in the gamma value is merely $0.0325$, the constraint $\|\gamma_i^s-\gamma_i^{s-1}\| \le \varepsilon_i$ in  Lemma \ref{lem:outer_basic} is naturally satisfied whenever $\varepsilon_i \ge 0.0325$. Therefore, this simulation sets the parameters $\delta$ and $\lambda$  to 0.05 and 1, respectively, taking into account the characteristic that parameter $\gamma$ is not suitable for frequent changes.

\begin{figure}
	\centering
	\begin{minipage}[h]{\linewidth}
        \subfigure[State distance without the proposed algorithm]{
			\includegraphics[width=0.46\linewidth]{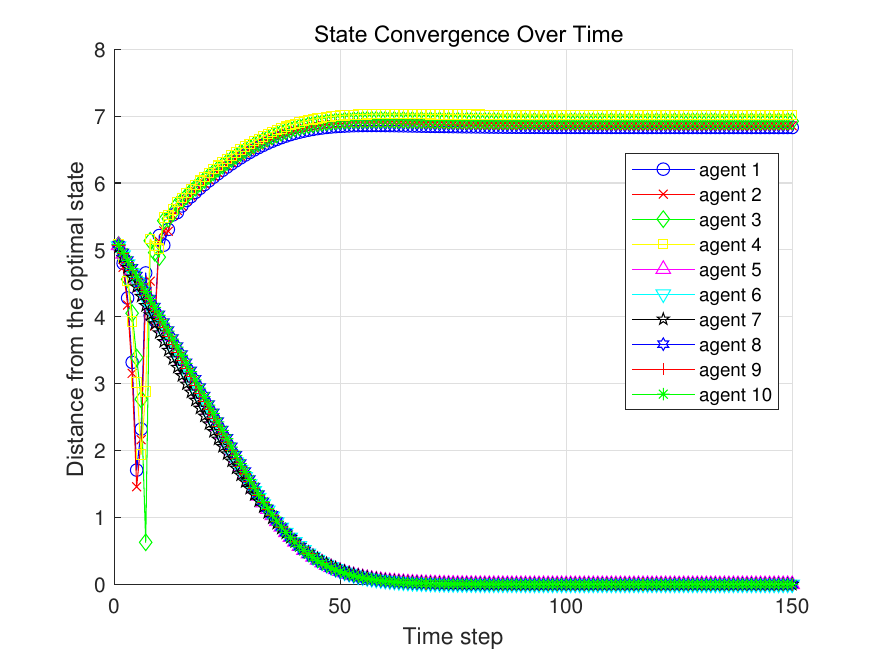}
		}
		\subfigure[Evolution cost  without the proposed algorithm ]{
			\includegraphics[width=0.46\linewidth]{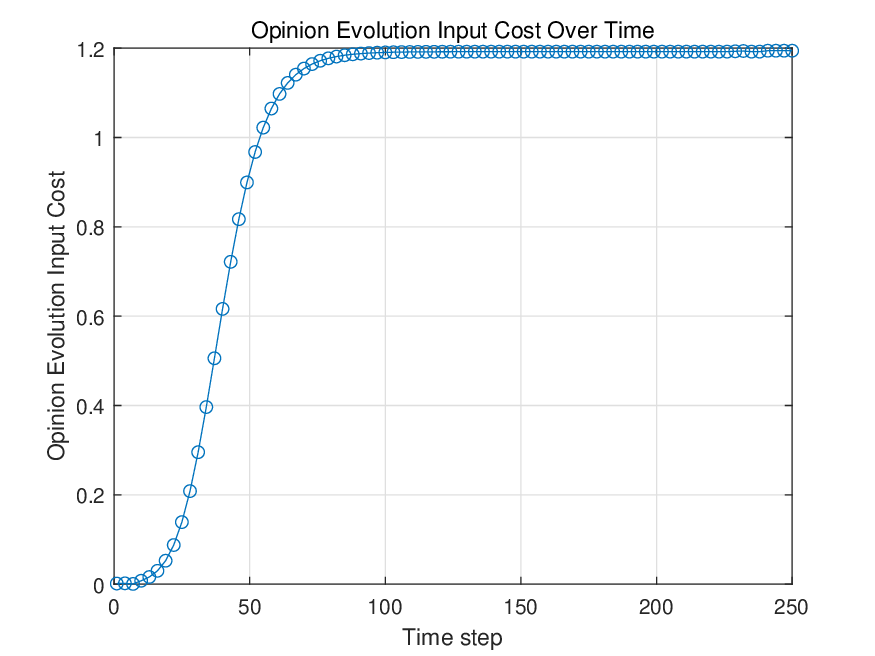}
		}
        \end{minipage}
		\begin{minipage}[h]{\linewidth}
        \subfigure[State distance with the proposed algorithm]{
			\includegraphics[width=0.46\linewidth]{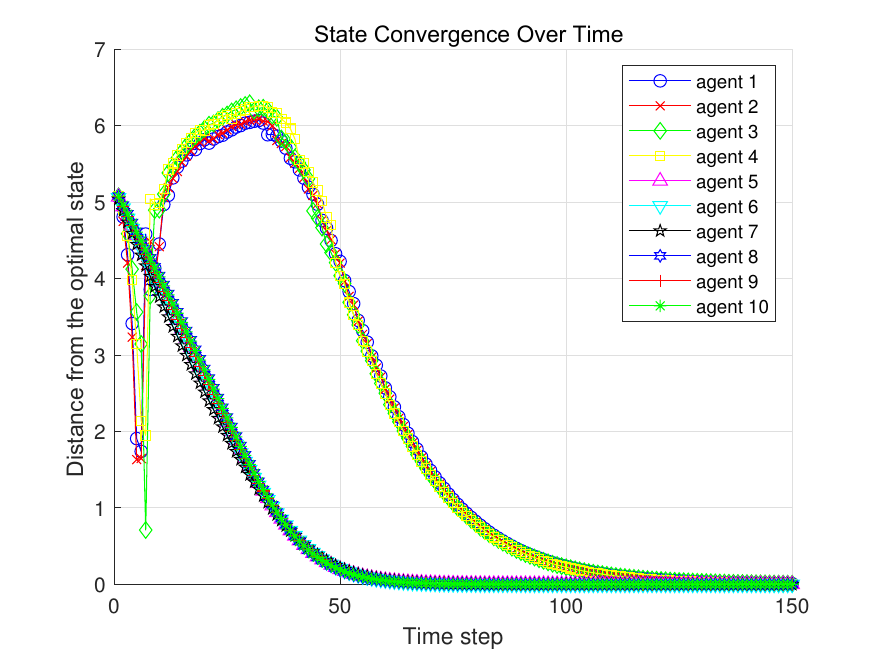}
		}
		\subfigure[Evolution  cost with the proposed algorithm]{
			\includegraphics[width=0.46\linewidth]{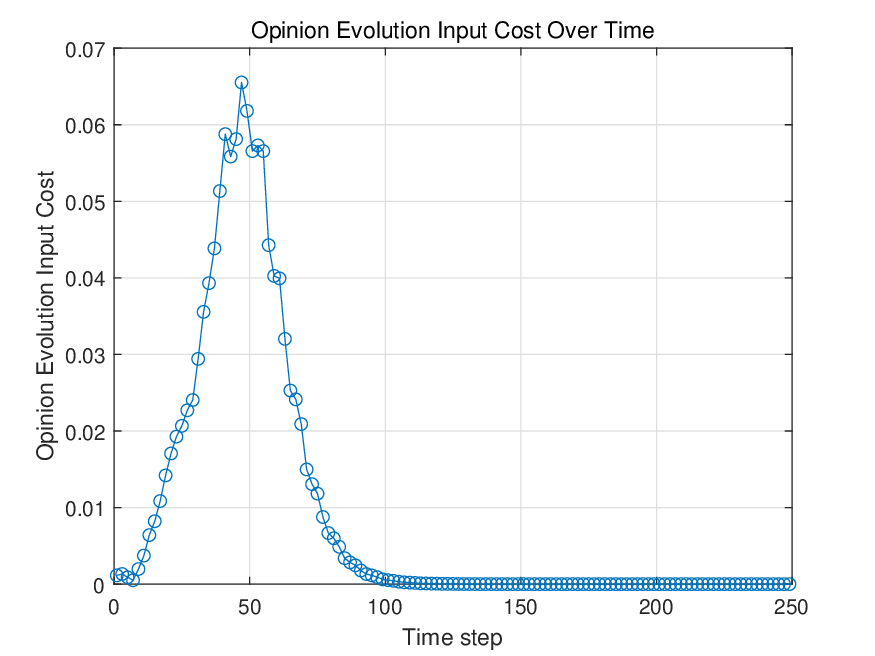}
		}
	\end{minipage}
	\caption{State distance and evolution cost with/without the proposed algorithm.}\label{opinion_dynamic_without_with_detect}
\end{figure}

First of all, in the absence of the proposed algorithm, simulation results show that normal agents’ states deviate from their targets and fail to converge, as illustrated in Fig.~\ref{opinion_dynamic_without_with_detect}(a). With a fixed $\gamma_i=0.1500$, malicious agents dominate the process, driving states away from social norms.

For comparison, Fig.~\ref{opinion_dynamic_without_with_detect}(c) shows that with the proposed mechanism, states evolve more slowly in the early stage, and the peak deviation occurs later than in Fig.~\ref{opinion_dynamic_without_with_detect}(a). This indicates that the proposed algorithm effectively reduces the early evolution speed, leaving room to further lower unnecessary costs.

Taking the first normal agent as an example, Fig.~\ref{opinion_dynamic_without_with_detect}(b)(d) compares the state evolution  cost with and without the proposed algorithm. Without isolation, the cost keeps rising and stabilizes at a much higher level, whereas with the algorithm it peaks early and then decreases as malicious links are removed. This shows that the proposed mechanism effectively reduces unnecessary long-term costs.

\begin{figure}
	\centering
	{
		\includegraphics[width=0.7\linewidth]{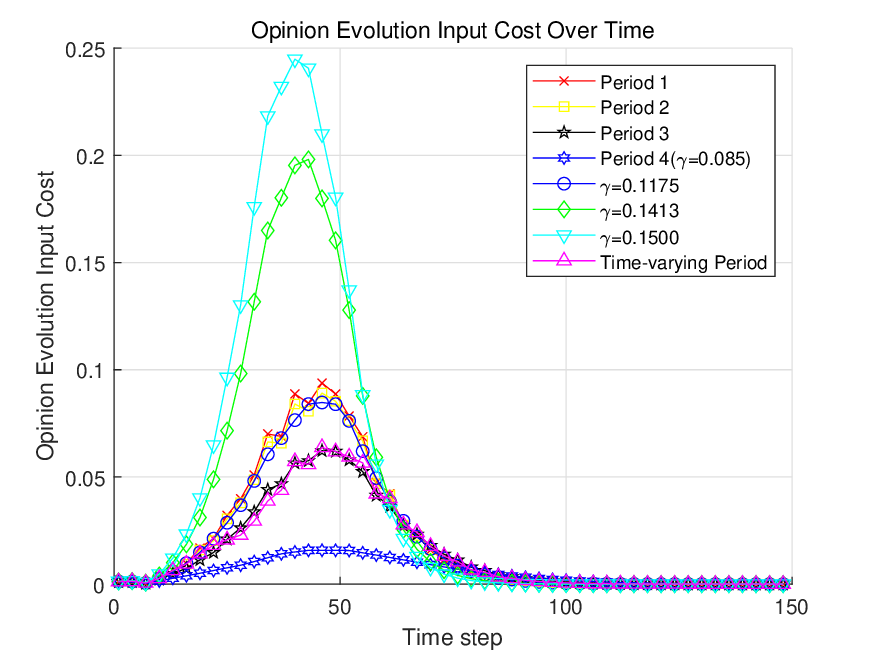}
	}
	\caption{State evolution cost with different function $f_i(k)$.}\label{u_vary_period}
\end{figure}

To further assess the proposed mechanism,  this simulation compares the norms of state evolution costs under various $f_i(k)$ functions. As shown in Fig.~\ref{u_vary_period}, lower peaks of $f_i(k)$ reduce control cost but slow convergence. The time-varying period function achieves a balance. It reduces costs during the early stages, thereby enabling the isolation of malicious agents. After the isolation is completed, it subsequently accelerates convergence.

Then, we introduce three evaluation metrics: early stage cost (ESC), later stage cost (LSC) and convergence step. 
Among them, ESC represents the area enclosed by the state evolution cost function and the coordinate axis until the peak of the state evolution cost; LSC represents the area enclosed by the state evolution cost function and the coordinate axis from the peak of the state evolution cost to convergence. And the convergence step is defined as the moment when the $g_i(k)R_ig_i(k)$ of input satisfies $g_i(k)R_ig_i(k) \le 0.001$. 

\begin{figure}
	\centering
	\begin{minipage}[h]{\linewidth}
		\subfigure[For Table I]{
			\includegraphics[width=0.45\linewidth]{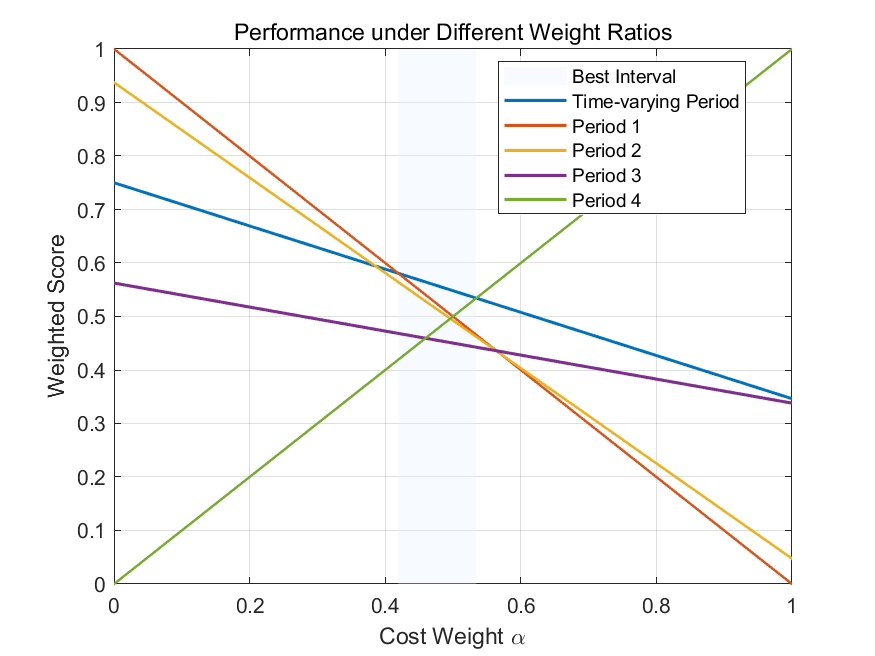}
		}
		\subfigure[For Table II]{
			\includegraphics[width=0.45\linewidth]{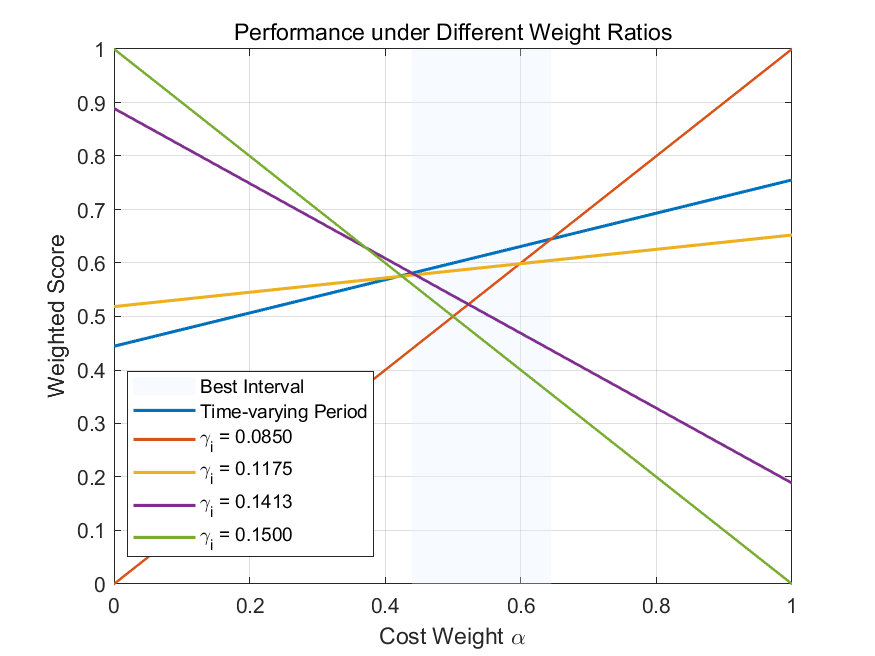}
		}
	\end{minipage}
	\caption{The best intervals for the time-varying period algorithm.}\label{best_interval}
\end{figure}

To demonstrate the effectiveness of the time-varying period algorithm from a quantitative perspective, this paper compares the proposed algorithm with different $f_i(k)$ functions.  As shown in Table \ref{tab:cost_convergence_step1}, the algorithm achieves a sound balance between cost and convergence steps. Furthermore, considering the total cost (ESC+LSC) and the convergence time as two separate objectives, the overall weighted score must be evaluated when the two objectives carry different weights. To this end, each objective is first normalized and then inverted so that a smaller original value corresponds to a larger normalized score. The normalization procedure follows $\hat{C}_i = 1 - \frac{C_i - \min(C)}{\max(C) - \min(C)}$ and $\hat{S}_i = 1 - \frac{S_i - \min(S)}{\max(S) - \min(S)}$,
where $C_i$ denotes the cost (ESC+LSC) and $S_i$ denotes the convergence step. After normalization, the overall score is computed by assigning a weight $\alpha$ to the cost and a weight $1-\alpha$ to the convergence step $ \text{Score}_i(\alpha)= \alpha\,\hat{C}_i + (1-\alpha)\,\hat{S}_i$.
By sweeping $\alpha$ from 0 to 1 with an increment of 0.01, the resulting scores are obtained and the best interval is plotted in Fig. \ref{best_interval}(a), which indicates that the corresponding best interval is $[0.4193,\, 0.5343]$.

In addition, the time-varying period algorithm is compared with a trust-based algorithm \cite{b06a1e49b8a44ea69478c5e207c0c9ed} with different constant $\gamma$. Table \ref{tab:cost_convergence_step2} shows that both algorithms maintain a favorable balance between cost and convergence steps. Fig. \ref{best_interval}(b) further shows that the best interval for the proposed algorithm is $[0.4396,\, 0.6449]$. 
The width of the best interval is used as an indicator of how well the algorithm balances cost and convergence steps. A wider interval centered around 0.5 reflects better stability and a stronger ability to achieve such balance. For the case in Table \ref{tab:cost_convergence_step1}, the interval width is 0.115. For the case in Table \ref{tab:cost_convergence_step2}, the interval width increases to 0.2053. These results demonstrate the effectiveness of the proposed algorithm. In particular, the results in Table \ref{tab:cost_convergence_step2} show that the proposed method outperforms and is more flexible than the approach based solely on trust values.

\begin{table}[h]
\renewcommand{\arraystretch}{1.3}
\centering
\caption{The evolution cost and convergence step under different $\gamma_i$ function $f_i(k)$}
\label{tab:cost_convergence_step1}
\begin{tabular}{l|ccc}
\hline
Function $f_i(k)$ & ESC & LSC & Convergence Step \\
\hline
\emph{Time-varying Period }& 1.1188 & 1.0853 & 97 \\
Period 1            & 1.7197 & 1.2240 & 93 \\
Period 2            & 1.6370 & 1.2054 & 94 \\
Period 3            & 1.1452 & 1.0765 & 100 \\
Period 4            & 0.3545 & 0.4536 & 109 \\
\hline
\end{tabular}
\end{table}

\begin{table}[h]
\renewcommand{\arraystretch}{1.3}
\centering
\caption{Comparative experiment with the trust-based algorithm in \cite{b06a1e49b8a44ea69478c5e207c0c9ed}}
\label{tab:cost_convergence_step2}
\begin{tabular}{l|ccc}
\hline
Function $f_i(k)$ & ESC & LSC & Convergence Step \\
\hline
\emph{Time-varying Period} & 1.1188 & 1.0853 & 97 \\
$\gamma_i=0.0850$   & 0.3545 & 0.4536 & 109 \\
$\gamma_i=0.1175$   & 1.5937 & 1.1982 & 95 \\
$\gamma_i=0.1413$   & 3.9314 & 1.5066 & 85 \\
$\gamma_i=0.1500$   & 5.0063 & 1.5075 & 82 \\
\hline
\end{tabular}
\end{table}

\section{Application of Strict Rejection Behavior in the Satellite-Assisted IoT Monitoring Network}

To validate the practicality of the proposed strict rejection behavior, a target capture scenario characterized by information uncertainty and malicious interference was constructed, wherein a group of functionally heterogeneous satellite-Assisted IoT devices is employed to monitor and track targets.


In this scenario, 2 types of satellites with different functionalities are considered. Both Type-A and Type-B satellites are capable of long-range coarse detection, and their observations can be characterized as lying within a circular region centered at the target. The key difference between them is that the observation accuracy of Type-A satellites is independent of distance, whereas that of Type-B satellites is distance-dependent: the closer the satellite is to the target, the smaller the observation error, and vice versa. In addition, only Type-A satellites are capable of executing the capture task.

Furthermore, 4 Type-B satellites are considered to observe the same space target, and their observations are transmitted to a central satellite (Type-A satellite) for information fusion and decision-making. The Type-A satellite performs control actions based on the fused target estimate to approach the target, while each Type-B satellite independently executes its own control based on its local observations. 

In this simulation, the Type-A satellite is designed to initiate the capture operation when the distance between its current position $p_A$ and the estimated target position $p_{t,A}^{\text{est}}$ is approximately $d_{\text{des}} = 80\mathrm{m}$. Therefore, the completion condition of the cooperative multi-satellite capture task is defined as
\begin{equation}\label{dist_thres}
\|\|p_A - p_{t,A}^{\text{est}}\|- d_{\text{des}}\|\leq \epsilon,
\end{equation}
where $\epsilon = 2\mathrm{m}$. In practical missions, the timing of the capture decision is of critical importance. The Type-A satellite is expected to make the capture decision as early as possible, in order to reduce the risk of collision caused by excessive proximity in the final approach phase\cite{Lamkin1991RPOCQFD}. By analogy with Lemma 1 in \cite{suo2025robust}, it can be obtained that, despite the presence of observation uncertainty during the satellite’s approach to the target, its state can still asymptotically reach a critical position, namely one that satisfies  (\ref{dist_thres}).

However, under the malicious interference considered in this simulation, one Type-B satellite is assumed to be compromised during information transmission and subjected to an injection attack as described in (\ref{malicious_inject}), resulting in its reported information deviating from the true value. Existing studies have shown that real-world satellite communication environments are indeed exposed to security threats such as malicious traffic injection, control-command tampering, and link-level attacks \cite{wan2026review,scholl2023introduction}. For example, attackers may inject malicious command traffic into telemetry, tracking, and command links, potentially causing abnormal satellite attitude, orbital deviation, or the implantation of backdoors \cite{idan2025aegissat}.
 Specifically, the malicious attacker deliberately injects an additional bias term $b_m$, i.e.,
\begin{equation}
p_{\text{mal}} = p_{t,B}^{\text{est}} + b_m \cdot \mathbf{d},
\end{equation}
which misleads the central satellite into believing that the target is located farther away than it actually is. As a consequence, the capture decision of the Type-A satellite is delayed, thereby reducing the overall mission success rate.
Here, $\mathbf{d}$ denotes the unit vector pointing away from the Type-A satellite. By adjusting the magnitude of $b_m$, different levels of malicious interference can be simulated, while varying the sign of $b_m$ allows modeling both normal fusion and adversarial fusion (i.e., fusion under strict rejection behavior). The fusion weights of the four Type-B satellites are initially set to $0.25$. Since this simulation focuses on the effect of strict rejection behavior, the weights are updated by decreasing $0.05$ every $50$ time steps. It is worth noting that, under this setting, the influence of the malicious node is never fully eliminated.
In addition, the parameter $\gamma$ for all satellites is initialized as $0.1$ and is updated every $50$ time steps.

In terms of dynamical modeling, the classical Clohessy–Wiltshire (C–W) relative motion model is adopted to describe the relative motion among satellites \cite{clohessy1960terminal}.

Let the orbital radius of the target be
$
R_e = 6.8781 \times 10^6 \text{m}.
$
Denote the relative position of the $i$-th satellite with respect to the target in the Hill frame as
$(x_{i,1}, x_{i,2}, x_{i,3})$, where $i \in \{0,1,2,3,4\}$.
Only in-plane motion is considered in this work, i.e., $x_{i,3} \equiv 0$.
Accordingly, the in-plane relative motion of satellite $i$ with respect to the target can be described as
\begin{equation}
\begin{aligned}
\ddot{x}_{i,1}
&=(\omega^2-\varrho_i\mu)(R_e+x_{i,1})
+2\omega \dot{x}_{i,2}
+a_{i,1}t, \\
\ddot{x}_{i,2}
&=-2\omega \dot{x}_{i,1}
+\omega^2 x_{i,2}
-\varrho_i \mu x_{i,2}
+a_{i,2}t.
\end{aligned}
\end{equation}
where
$
\varrho_i = \Big((R_e+x_{i,1})^2+x_{i,2}^2\Big)^{{-}3/2}$, $\mu = 3.9860\times 10^{14}\ \text{m}^3/\text{s}^2$ denotes the Earth's gravitational constant, $\omega = \left(\mu/{R_e^3}\right)^{1/2}$
 represents the mean orbital angular velocity, and $(a_{i,1}, a_{i,2})$ denote the additional acceleration inputs caused by aerodynamic drag.

By linearizing the above nonlinear model, the model can be obtained as $\dot{\bar{x}}_i(t)=\bar{A}\bar{x}_i(t)+\bar{B}\bar{u}_i t$, $
y_i(t)=\bar{C}\bar{x}_i(t)$, $
\quad i\in\{1,2,3,4,5\}$.
The state and input vectors are defined as $x_i = [p_{i,x}, p_{i,y}, v_{i,x}, v_{i,y}]^T$, representing the relative position and velocity states, and $u = [a_{i,1}, a_{i,2}]^T$ denotes the control input.
The system matrices are denoted by
$\bar{A}=[0, I_2; \bar{A}_{21}, \bar{A}_{22}]$,
$\bar{B}=[0; I_2]$, and
$\bar{C}=[I_2, 0]$, where
$\bar{A}_{21}=[3\omega^2, 0; 0, 0]$ and
$\bar{A}_{22}=[0, 2\omega; -2\omega, 0]$.
\begin{figure*}
	\centering
	\begin{minipage}[!htb]{\linewidth}
		\subfigure[Time evolution of the distance to the true target]{
			\includegraphics[width=0.33\linewidth]{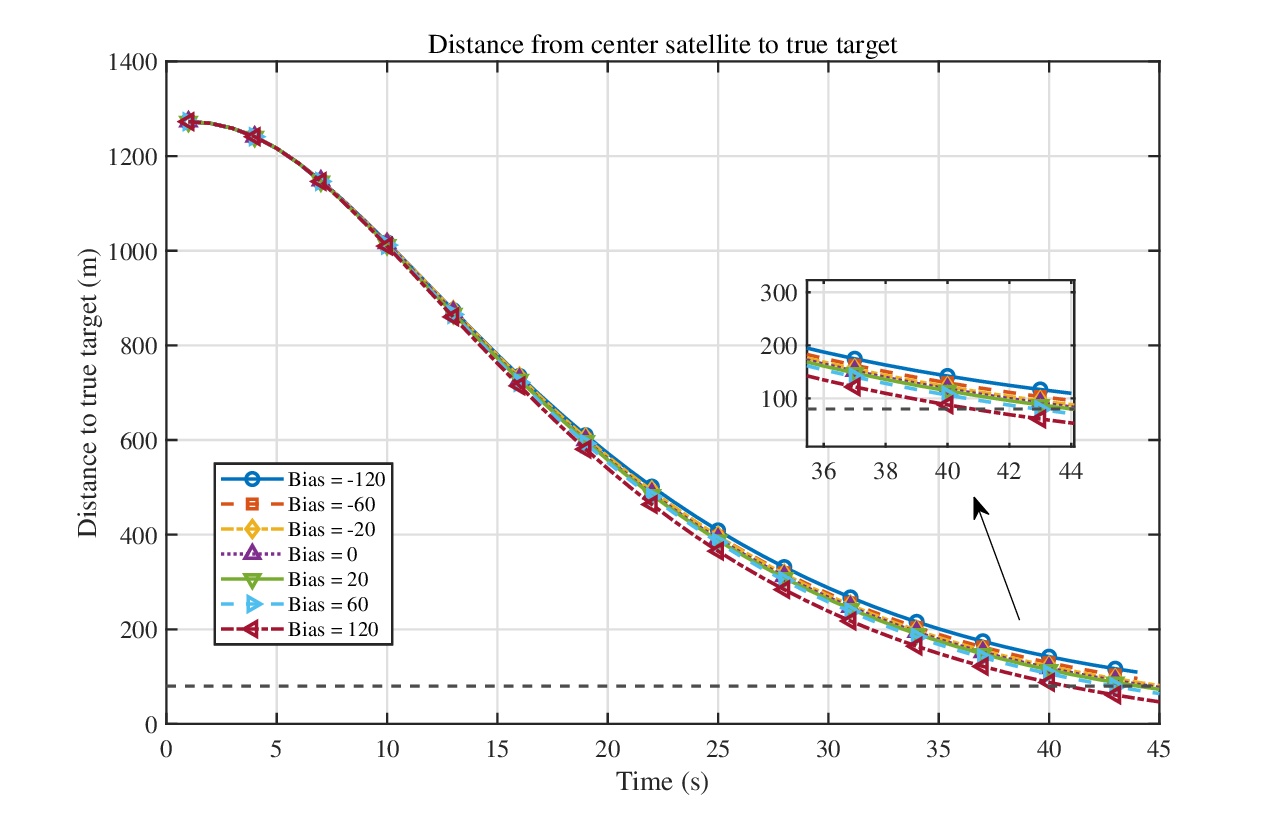}
		}
		\subfigure[Final distances under different malicious biases]{
			\includegraphics[width=0.33\linewidth]{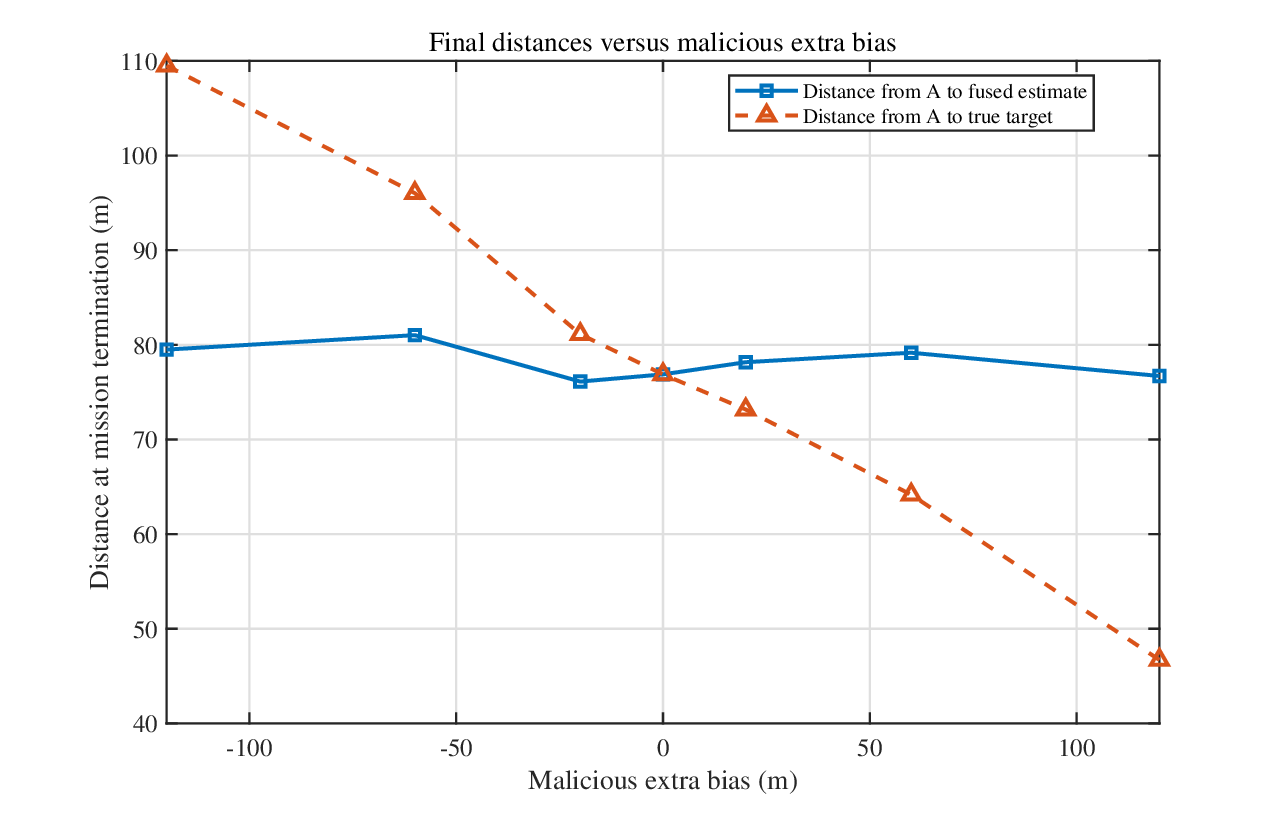}
		}
        \subfigure[Control cost under different malicious biases]{
			\includegraphics[width=0.33\linewidth]{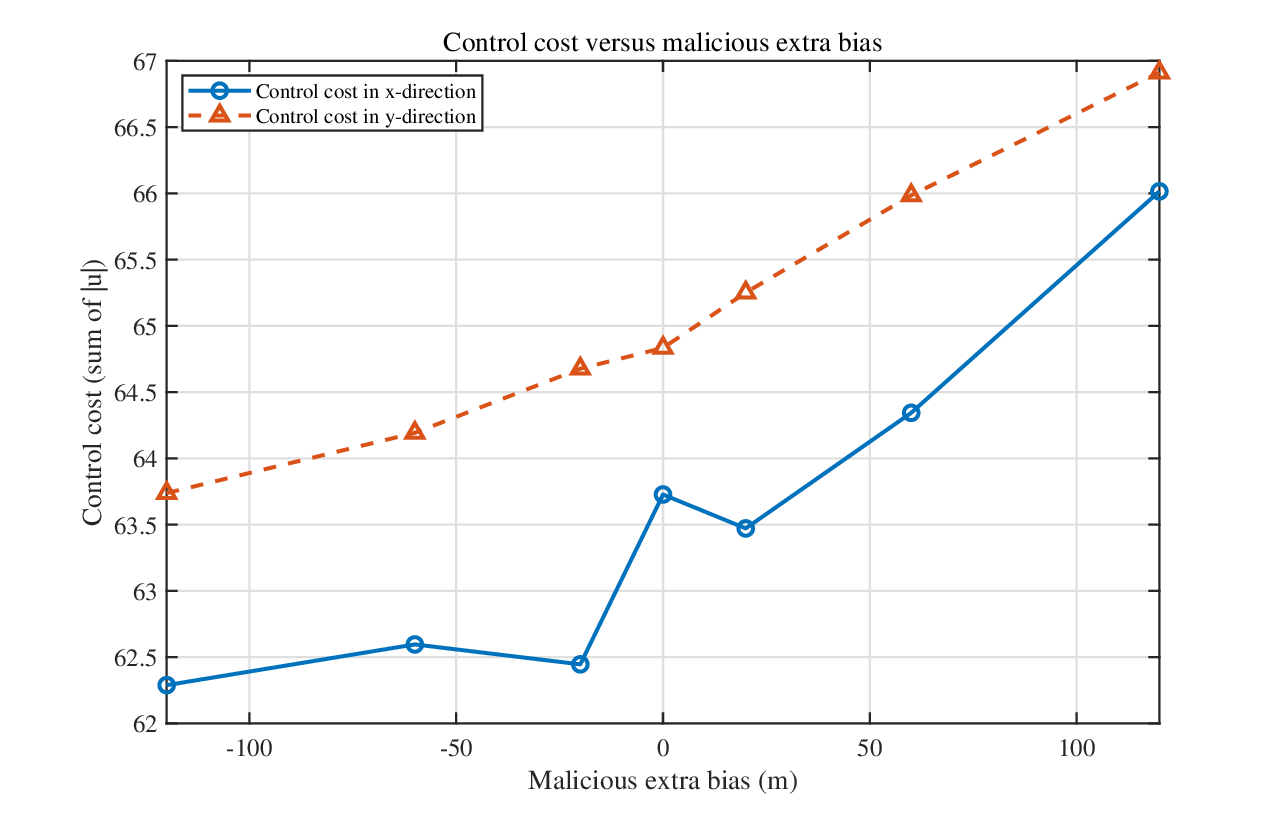}
		}
	\end{minipage}
	\caption{Performance comparison under different malicious extra biases.}
	\label{distance_fig}
\end{figure*}
Then, the continuous-time system is discretized as $x_i(k+1)=Ax_i(k)+Bu_i(k)$,
where the discrete-time system matrices are given by
$
A=e^{\bar{A}T_s}$, $
B=\int_0^{T_s} e^{\bar{A}\tau}\, d\tau\, \bar{B}$, $
C=\bar{C}$,
and the sampling period is chosen as $T_s=1$. Moreover, it is assumed that the distances among all Type-B satellites are much smaller than the sensing range. Therefore, the entire formation can be approximated as a point mass. The initial states of the 5 satellites are set as $x(0)=
[
-900, -900, 0, 0
]^{T}$. Based on the above modeling framework, the impact of the fusion process on task performance can be systematically analyzed under scenarios with and without strict rejection behavior.

First, the task completion performance is evaluated. As shown in Fig.~\ref{distance_fig}(a), the distance between the central satellite and the true target position over the entire capture process is presented. The seven curves correspond to the cases with $b_m = [-120, -60, -20, 0, 20, 60, 120]$, respectively. It can be observed that, in all cases, the satellites are able to approach the target and successfully trigger the capture decision.

Furthermore, as illustrated in Fig.~\ref{distance_fig}(b), both the normal fusion scheme and the fusion scheme with Strict Rejection Behavior satisfy the requirement in (\ref{dist_thres}) at the moment of capture decision. However, under normal fusion (i.e., without Strict Rejection Behavior), the actual distance between the central satellite and the true target position at the capture moment is often smaller than $80\mathrm{m}$. This indicates that malicious information leads to delayed capture decisions, causing the satellite to get excessively close to the target and thus increasing the risk of collision. In contrast, when Strict Rejection Behavior is incorporated, the central satellite is able to make the capture decision earlier, before reaching the $80\mathrm{m}$ threshold, thereby significantly improving operational safety.

Next, Fig.~\ref{distance_fig}(c) presents the accumulated magnitude of control inputs over the entire time horizon for each case. It can be observed that, under normal fusion, the control cost increases as the level of malicious interference becomes stronger. In contrast, under the proposed strict rejection behavior  fusion mechanism, the control cost decreases as the malicious interference intensifies. This phenomenon can be attributed to the positive correlation between the magnitude of control inputs and the distance between the central satellite and the estimated target position.

\begin{rem}
    It should be noted that the effectiveness of strict rejection behavior is highly dependent on the specific task scenario. For example, in Section 4, the attacker primarily aims to slow down the process by which normal agents approach the target, whereas in this section, the objective is to delay the timing of capture decisions made by normal agents. Therefore, strict rejection behavior should be designed and adjusted in a task-specific manner.
\end{rem}

\section{Conclusions}

This paper studied cost-aware distributed online learning in IoT-enabled MASs under persistent adversarial interactions. A strict rejection behavior was introduced to suppress the harmful assimilation of suspicious neighboring information. It was further shown that such rejection may induce heterogeneous transient evolution among neighboring normal agents, leading to evolution desynchronization across the network. To mitigate this effect, a two-time-scale adaptive evolution regulation framework was developed, and its dynamic tracking property and propagation attenuation capability were theoretically established. Numerical simulations and a satellite-assisted IoT monitoring scenario verified the effectiveness and practicality of the proposed method. Future work will focus on extending the proposed framework to more
general application scenarios, particularly those involving varying information structures.

\appendices

\section{An Example of the Reference schedule generation}
\medskip
 Let
$
\sigma_i^s \in \{0,1\}
$
be a binary indicator available at the beginning of the $s$-th outer period, where $\sigma_i^s=1$ means that agent $i$ is under suspected malicious influence, and $\sigma_i^s=0$ otherwise. 
Let
$
\mathcal{F}_i:=\{f_{i,1},\ldots,f_{i,L}\}$, $
f_{i,\min}=f_{i,L}<\cdots<f_{i,1}=f_{i,\max},
$
be a set of admissible quantization levels. Given the current periodic schedule $\gamma_i^{s-1}$,   a reference schedule
$
\gamma_{i,\mathrm{ref}}^s
:=\mathrm{col}\!\big(\gamma_{i,\mathrm{ref}}^{(1,s)},\ldots,\gamma_{i,\mathrm{ref}}^{(T,s)}\big)\in\mathbb{R}^T
$ can be constructed
by modifying its peak entries only. More precisely, let
$
\gamma_{i,\max}^{s-1}:=\max_{\tau\in\{1,\ldots,T\}}\gamma_i^{(\tau,s-1)}
$.
Then,  
 a specific example of warning-responsive peak clipping is described Algorithm~\ref{alg:peak_clipping}.

\begin{algorithm}
\caption{Reference schedule generation by peak clipping}
\label{alg:peak_clipping}
\begin{algorithmic}[1]
\Require Current schedule $\gamma_i^{s-1}=\mathrm{col}(\gamma_i^{(1,s-1)},\ldots,\gamma_i^{(T,s-1)})$, quantization levels $\mathcal{F}_i=\{f_{i,\ell}\}_{\ell=1}^L$, warning signal $\sigma_i^s\in\{0,1\}$
\Ensure Reference schedule $\gamma_{i,\mathrm{ref}}^s$
\State Compute $\gamma_{i,\max}^{s-1}=\max_{\tau}\gamma_i^{(\tau,s-1)}$.
\State Find two adjacent levels $f_{i,\ell_1}$ and $f_{i,\ell_2}$ such that
$
f_{i,\ell_1}\le \gamma_{i,\max}^{s-1}\le f_{i,\ell_2}
$
with $f_{i,\ell_1}$ the nearest lower level and $f_{i,\ell_2}$ the nearest upper level.
\State Initialize $\gamma_{i,\mathrm{ref}}^s \gets \gamma_i^{s-1}$.
\If{$\sigma_i^s=1$}
    \State Replace every entry equal to $\gamma_{i,\max}^{s-1}$ in $\gamma_{i,\mathrm{ref}}^s$ by $f_{i,\ell_1}$.
\Else
    \State Replace every entry equal to $\gamma_{i,\max}^{s-1}$ in $\gamma_{i,\mathrm{ref}}^s$ by $f_{i,\ell_2}$.
\EndIf
\State \Return $\gamma_{i,\mathrm{ref}}^s$.
\end{algorithmic}
\end{algorithm}

\bibliographystyle{IEEEtran}
\bibliography{ref}











\newpage

 




\vfill

\end{document}